\documentclass[showpacs,twocolumn,floatfix]{revtex4-1}

\usepackage{graphicx,psfrag,amsmath,amssymb,amsfonts,latexsym,color,dcolumn,epsf,graphpap}
\begin{document}

\title{Correction to the geometric phase by structured environments: the onset of non-Markovian effects}
\author{Fernando C. Lombardo and Paula I. Villar}
\affiliation{Departamento de F\'\i sica {\it Juan Jos\'e
Giambiagi}, FCEyN UBA and IFIBA CONICET-UBA, Facultad de Ciencias Exactas y Naturales,
Ciudad Universitaria, Pabell\' on I, 1428 Buenos Aires, Argentina}

\date{today}

\begin{abstract}
We study the geometric phase of a two-level system under the presence of a structured environment, particularly analysing its correction with the ohmicity parameter $s$ and the onset of non-Markovianity. We firstly examine the system coupled to a set of harmonic oscillators and studied the decoherence factor as function of the environment's ohmicity parameter. Secondly, we propose the two-level system coupled to a non-equilibrium environment, and show that these environments display non-Markovian effects for all values of the ohmicity parameter. 
The geometric phase of the two-level system is therefore computed under the presence of both types of environment. The correction to the unitary geometric phase is  analysed in both, Markovian and non-Markovian regimes. Under Markovian environments, the correction induced 
on the system's phase is mainly ruled by the coupling constant between the system and the environment, while in the non-Markovian regime, memory effects seem to  trigger a significant correction to the unitary geometric phase. The result is significant to the quantum information
processing based on the geometric phase in quantum open systems.
\end{abstract}
\maketitle

\section{Introduction}

A system can retain the information of
its motion when it undergoes a cyclic evolution, in the form of a
geometric phase, which was first put forward by Pancharatnam in
optics \cite{Pancharatman} and later studied explicitly by Berry in
a general quantal system \cite{Berry}. Since then, great progress
has been achieved in this field.  For example, the application of the geometric phase has
been proposed in many fields, such as the geometric quantum
computation.  In this
line of work, many physical systems  have  been  investigated  to
realise  geometric  quantum  computation,  such  as  NMR  (Nuclear
Magnetic Resonance) \cite{NMR}, Josephson  junction \cite{JJ},  Ion  trap \cite{IT}  and
semiconductor  quantum  dots \cite{QD}. 
The quantum computation scheme for the 
geometric phase has been proposed based on the Abelian  or
non-Abelian geometric concepts, and the geometric phase has been shown
to be robust against faults in the presence of some kind of
external noise due to the geometric nature of Berry phase \cite{refs1, refs2, refs3,xu}. 
Then, for isolated quantum systems, the geometric phase is theoretically perfectly understood and 
experimentally verified. However, it has been shown that the interactions play an important role for the
realisation of some specific operations. As the gates operate slowly compared to the 
dynamical time scale, they become vulnerable to open system effects and parameters' fluctuations
that may lead to a loss of coherence. Consequently, the study of the
geometric phase was soon extended to open quantum systems. Following this idea,
 many authors have analysed the correction to the geometric phase
 under the influence of an external thermal or non-equilibrium environments, using different
approaches (see \cite{Tong, Gefen, pra, nos, pau, pra87}). 
In all cases, the purely dephasing model considered was a spin-$1/2$ particle coupled to the
environment's degrees of freedom through a $\sigma_z$-coupling.
The interest on the geometric phase in open systems has also been extended to some experimental setups
\cite{prl}. 
Lately, it has also been observed in a variety of superconducting 
systems \cite{ scienceexp, berger}, and it has been shown the importance of quantifying decoherence 
when geometric operations are carried out in the presence of low-frequency noise.
All real experiments generally imply the presence of an external environment which induces noise and dissipation on the subsystem depending on the strength of the coupling among them.
Memory effects are also considered a relevant source of noise that can affect the dynamics of the
system of interest.
Thus, a detailed mechanism of decoherence due to external noise sources is still required in order to overcome the effect of a destructive decoherence in measurements of geometric phase.

Within a microscopic approach, quantum Markovian master equations are usually obtained by means of the Born-Markov approximation, which assumes a weak system-environment coupling and several further, mostly rather drastic approximations \cite{Breuer}.  However, these approximations are not applicable  in many processes occurring in nature, such as strong environment couplings, structured and finite reservoirs, low temperatures and large initial system-environment correlations \cite{Breuerlibro}. In the case of any substantial deviation from the dynamics of a quantum Markov process, one often speaks of a non-Markovian process, implying that the dynamics is governed by significant memory effects. Unfortunately, a consistent general theory of quantum non-Markovianity does not exist and even the very definition of non-Markovianity is a highly up-to-date issue. Very recently important steps towards a general theory of non-Markovianity have been made. There has been a great effort  to rigorously define the border between Markovian and non-Markovian quantum dynamics and to develop quantitative measures for the degree of memory effects in open systems \cite{Wolf,Breuerrev,Rivas}. The quantification of non-Markovianity is justified by the fact that there is
increasing evidence of its crucial role as a resource for quantum technologies \cite{Vasile,Rivas}. Non-Markovianity evolution is often characterised by decoherence phenomena and information trapping, hence
leading to longer coherence times in comparison to the Markovian case.
As  reservoir engineering techniques become
experimentally feasible, it is crucial to establish relations between the occurrence of non-Markovianity
and the form of the environmental spectrum. Recently, there has been many studies on systems whose
reduced dynamics are characterised by memory effects.
Non-Markovian features play an important role in systems where the frequency spectrum of the environment is
structured. In Ref.\cite{haikka}, authors established a connection between the general form of the spectrum and
the memory effects in the reduced system dynamics.

In this framework,  this study has been two-folded motivated: on the one side, the need to have a better understanding of the geometric phase in open quantum systems in order  to achieve fault tolerance quantum computation.
 On the other side,  the need to understand memory effects as a source of noise for the system of interest. In this manuscript, we shall study the correction to the geometric phase of a two-level system coupled to an external environment.  
We shall use the spin-boson model since it has the advantage of having an analytical solution which can be useful to have a better insight into the onset of non-Markovian effects. By the use of the established relation
between the form of the reservoir spectrum and the onset of non-Markovianity, we shall study the 
geometric phase of the open system as a function of the ``ohmicity'' of the reservoir, which allows the description
of sub-Ohmic, Ohmic and super-Ohmic spectrum. The paper is organised as follows: in Section II,
we briefly described the model and present both types of environments to be analysed: thermal equilibrium and non-equilibrium ones. We study the diffusion coefficients and the corresponding decoherence factors. Following Ref.\cite{haikka} we study the onset of non-Markovianity in the environments considered. In Section III, we compute the geometric phase acquired by the system for 
different forms of the reservoir spectrum, as function of the ohmicity parameter and study how the memory effects of the environment affect the geometric phase. Finally, in Section IV, we shall make our final remarks.

\section{The model}

The spin-boson model is studied in a variety of fields, such as
condensed matter physics, quantum optics, physical chemistry and
quantum information science \cite{legget} in order to describe
non-unitary effects induced in quantum systems due to a
coupling with an external environment.
For a quantum system, the influence of the surroundings 
plays a role at a fundamental level. When the environment 
is taken into consideration, the system dynamics can no 
longer be described in terms of pure quantum
states and unitary evolution. From a practical point of view,
all real systems interact with an environment,
which means that we expect their quantum evolution to be plagued by non-unitary
effects, namely dissipation and decoherence. 
Most theoretical investigations of how the system is affected
by the presence of an environment have been done using a thermal reservoir,
usually assuming Markovian statistical properties and defined bath correlations 
\cite{paz-zurek, weiss} (there are also works on non-Markovian models as, just for example, \cite{nonMark}). 

In the following, we shall consider a paradigmatic spin-boson model consisting 
of a two-level system coupled to an external environment. We shall consider two different types of environments and see the non-Markovian effects:  thermal equilibrium and non-equilibrium environment. In both cases, we shall compute the diffusion and decoherence factor derived by the definition of the corresponding bath correlations and the spectral density $I(\omega)$, which can be defined for a general environment as:
\begin{equation}
 I(\omega) = \gamma_0 \frac{\omega^s}{\Lambda^{s-1}}\exp(-\omega/\Lambda).
 \label{densidad}
\end{equation}
In Eq.(\ref{densidad}), $\Lambda$ is the reservoir frequency cutoff and $\gamma_0$ is the coupling constant (has different units for the different environment considered). By changing the value of the $s$-parameter
one goes from sub-Ohmic reservoirs ($s<1$) to Ohmic ($s=1$) and super-Ohmic ($s>1$) reservoirs, respectively.
The Ohmic spectrum is generally used to describe charged interstitials in metals. The supra-Ohmic environment
commonly describes the effect of the interaction between a charged particle and its electromagnetic field.
The sub-ohmic environment is used to model the type of noise occurring in solid state devices due to low frequency modes,
similar to the ``1/f'' noise in Josephson junctions. Thus, the description of the spectral density function in terms of
a continuous parameter $s$ allows to simulate paradigmatic models of open quantum systems by the variation of $s$.
We stress that such engineering of the ohmicity of the spectrum is possible when simulating the dephasing
model in trapped ultracold atoms \cite{haikka2}.

\subsection{Thermal equilibrium environments and the onset of non-Markovianity}

We shall start by studying the decoherence effects induced by a thermal equilibrium environment, generally modelled by a set of harmonic oscillators.
The interaction between the two-state system and the environment
is entirely represented by a Hamiltonian in which the coupling is
only through $\sigma_z$. In this particular case,
$[\sigma_z,H_{\rm int}]=0$ and the corresponding master equation for the reduced density matrix, is
much simplified, with neither frequency renormalisation nor dissipation
effects. The Hamiltonian of
the complete system is written as:

\begin{equation}
H_{\rm SB}=  \frac{1}{2} \hbar \Omega
 \sigma_z +\frac{1}{2} \sigma_z
\sum_{k} \lambda_{k} (g_k a^{\dagger}_{k}+g_k^*a_{k}) + 
\sum_{k} \hbar \omega_{k}
a_{k}^{\dag} a_{k}. \label{HSB}
\end{equation}
The interaction hamiltonian is only proportional to $\sigma_z$ which means that the populations of the
eigenstates remain constant while the off-diagonal terms of the reduced density matrix decay 
due to the existence of the environment, as
\begin{equation}
 \rho_{r_{10}}(t)=\rho_{r_{01}}^*(t)=\rho_{r_{10}}(0) e^{-{\cal F}(t)} e^{-i \Omega t},
\end{equation}
where  $\cal F$ is called the decoherence factor defined as (see Ref.\cite{pra} for details)
\begin{eqnarray}
{\cal F}(t) &=& 2 \int_0^{\infty} d\omega I(\omega) \coth\bigg(\frac{\omega}{2 k_B T}\bigg)
\frac{(1-\cos(\omega t))}{\omega^2} \nonumber \\
&=& 2 \int_0^{t} ds D(s),
\label{decof}
\end{eqnarray}
where $D(s)$ is the diffusion coefficient, present in the master equation.
It is clear that ${\cal F}(t)$ not only depends on the temperature of the bath but also on the spectral density, particularly on the ohmicity parameter $s$.\\

In Ref.\cite{haikka}, authors have shown  that there exists a link between the onset of non-Markovianity and the form
of the reservoir spectrum for thermal equilibrium environments as the ones considered here. They have stated that the crossover is signalled by the onset of periods during which the diffusion rate 
is negative. A common feature of all non-Markovianity measures is that they are based on the non-monotonic evolution in time of certain 
coefficients which signal the information back-flow from the environment back to the main system. Without memory effects, decoherence 
rates (as a quantum channel) decreases in time monotonically. However, environment memory effects may produce a non-monotonic behaviour 
of the quantum channel capacity. This is related to the convex to non-convex changes of the spectrum. The condition on the non-convexity of the environmental spectrum is a necessary and sufficient condition for non-Markovianity at all temperatures. This has been set as $s > 2$ for zero-T environments. Hence, memory effects leading to information back flow and
recoherence occur only if the reservoir spectrum is super-Ohmic with $s>2$. This means that even if the reduced
dynamics is exact, and hence no Markovian approximation has been performed, the time evolution of the qubit
does not present any memory effects for Ohmic and sub-Ohmic spectra. This argument has been derived based
on the non-monotonic behaviour of the decoherence factor ${\cal F}(t)$, and basically it is equivalent to the fact that 
the diffusion coefficient $D(s)$ becomes negative.

\begin{figure}[!ht]
\includegraphics[width=8.5cm]{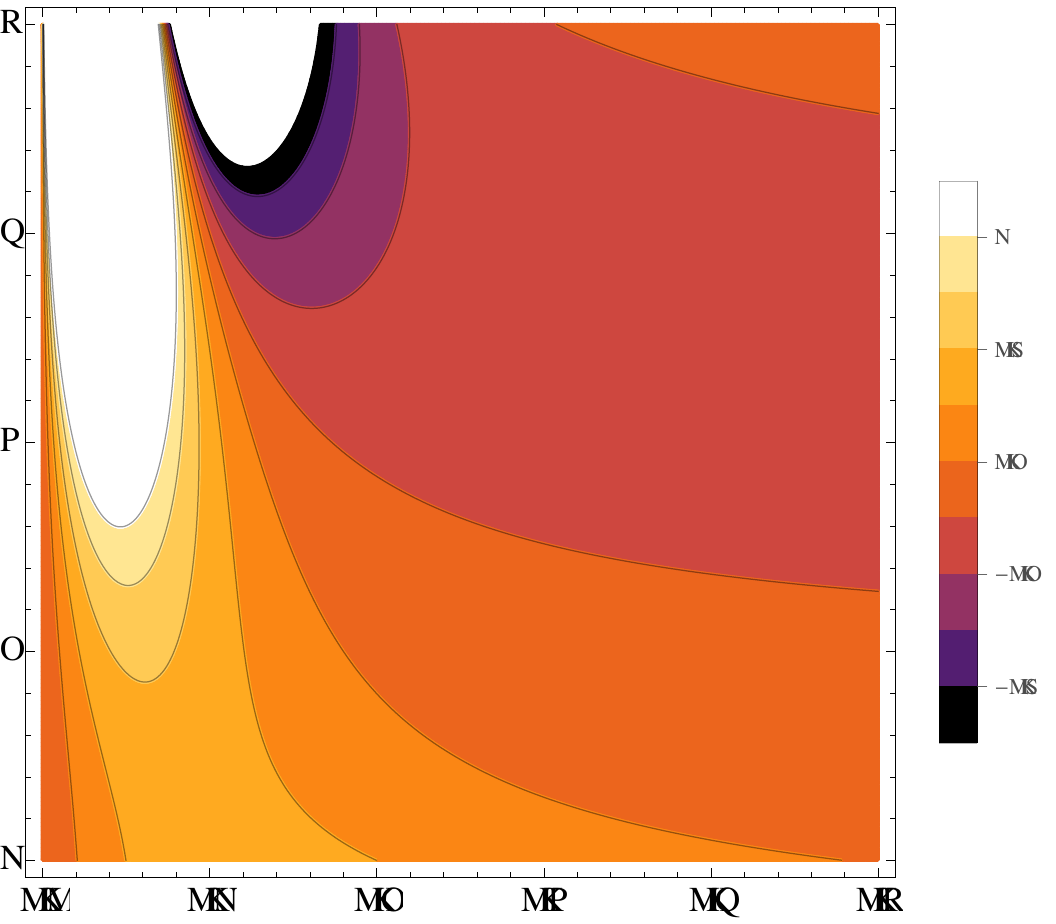}
\caption{(Color online) Density plot of the diffusion coefficient of thermal environments. 
The horizontal axis represents the time evolution as $\Omega t$ and the vertical axis is the ohmicity $s$. The darkest areas are negative values of the diffusion coefficient which has been shown to be related to the onset of non-Markovian effects as stated in \cite{haikka}. Parameters used: $\gamma_0= 0.1$ and $\Lambda=10 \Omega$.}
\label{dif}
\end{figure}

In Fig.\ref{dif}, we present a contour plot for  the diffusion coefficient for different equilibrium environments at zero temperature, to show the onset of non-Markovianity 
as derived in \cite{haikka}. We can see the regions where the diffusion coefficient has a negative value, which are related to the 
onset of non-Markovian effects. For example, in Fig.\ref{dif}, we see that for Ohmic thermal environments, i.e. $s=1$ in the vertical axis, we always have positive values for the diffusion coefficient.
This means that the Ohmic environment is always Markovian. For other environment, such as $s=4$ in the vertical axis, we can find regions where the diffusion coefficient gets positive values and areas where the value is negative. This means that the memory effects become important after some time evolution.\\

Using Eqs.  (\ref{densidad}) and (\ref{decof}), the 
exact decoherence factor as function of time and the ohmicity parameter $s$, can be written as:
\begin{eqnarray}
{\cal F}(t) &=& 4\gamma_0 \frac{\Gamma[s]}{s - 1} \left[1 - \left(1 + \Lambda^2t^2\right)^{-\frac{s}{2}}\right.
\nonumber \\
&\times & \left. \left(\cos[s \arctan(\Lambda t)] + \Lambda t \sin[s \arctan(\Lambda t)] \right)\right],
\label{factorexacto}
\end{eqnarray}
where $\Gamma[x]$ is the Gamma function.
\begin{figure}[!ht]
\includegraphics[width=8.5cm]{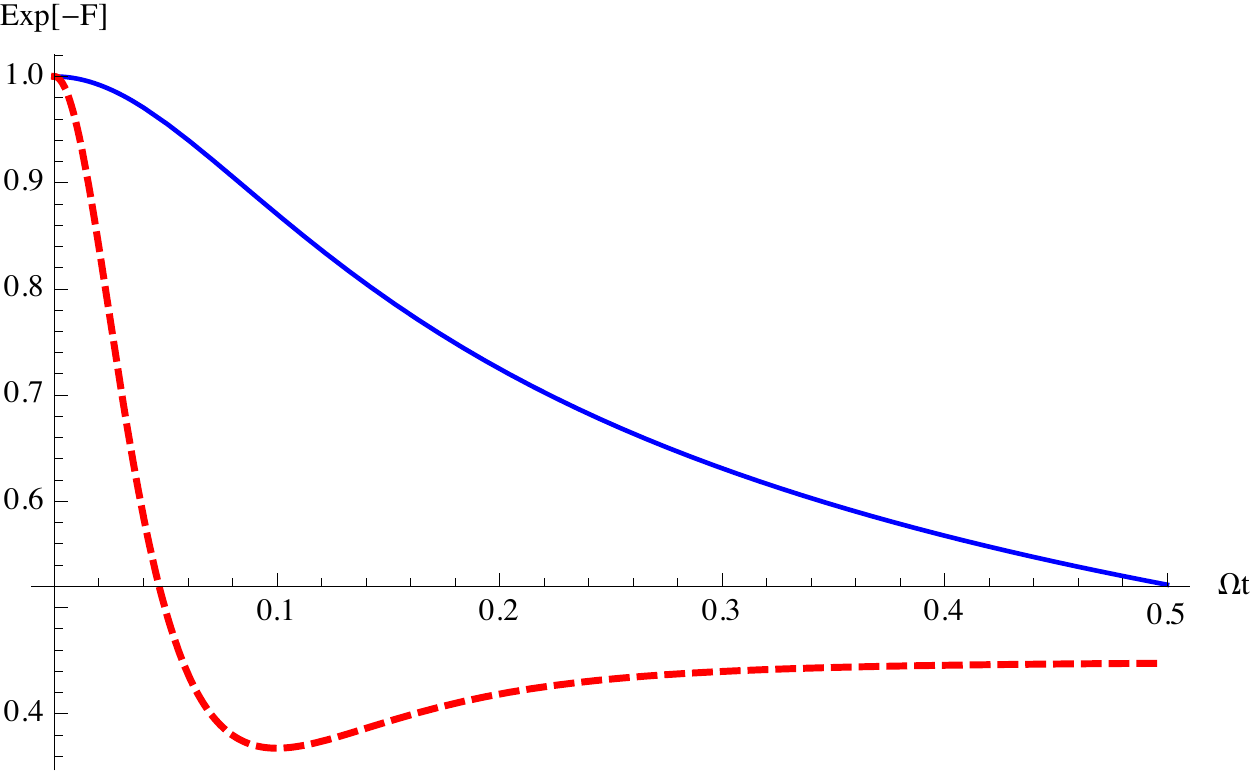}
\caption{(Color online) Decoherence factor as function of the time for different values of the ohmicity parameter $s$.  It can be seen that areas with negative diffusion coefficient in Fig.\ref{dif} present a non-monotonic behaviour in the decoherence factor leading to recoherence effects.
The  blue line corresponds to $s=1$ and the red dashed one corresponds to $s=4$. Parameters used: $\gamma_0= 0.1$ and $\Lambda=10 \Omega$.}
\label{factorequi}
\end{figure}
In Fig.\ref{factorequi}, we present the evolution of the decoherence factor as time evolves.
In Ohmic ($s=1$) thermal environments, the decoherence factor is known to be a monotonic decreasing function in time (blue line in Fig.\ref{factorequi}). However, when the  memory effects of the environment become relevant (for example $s=4$), the decoherence function has not a monotonic behaviour, leading to recoherence phenomena (red dashed line in Fig.\ref{factorequi}). This is related to the negative regions of the diffusion coefficient as can be seen in Fig.\ref{dif}.

\subsection{Non-equilibrium environments and the onset of non-Markovianity}

In this subsection, we shall deal with non-equilibrium environments, represented
by random perturbations with non-stationary statistics. The motivation to study these type of environments is twofold: on one side, the diffusion coefficients
computed for different environment spectra have negative parts for all values of the ohmicity parameter. This means they can 
be considered as non-Markovian environments for all values of the parameter $s$ and, contrary to the  environments studied above,  there is not a critical value of $s$ which determines the onset of non-Markovianity. On the other side, these non-equilibrium baths may represent a proposal for engineering
reservoirs in a manner reminiscent of a coherent control experiment using shaped pulses \cite{pulses}.  In this 
model, the control parameter $\lambda$, is derived not from a laser pulse, but rather from well-defined phase 
relations between the modes of the bath.

The modelling of these non-equilibrium environments implies that the two-level quantum system presents an energy gap 
$E_2(t)-E_1(t)= \hbar \omega(t)$ which fluctuates due to the influence of the environment,
where $E_j(t)$, with $j=1,2$ is the instantaneous energy of state $j$ as perturbed
by the surroundings. Following the idea
proposed in \cite{martenschem}, the bath is represented by a random function
of time corresponding to the transition frequency of the two-state quantum
system. In contrast to the usual treatment, the statistical properties
of this random function are non-stationary, corresponding physically to, for example,
impulsively excited phonons of the environment with initial phases that are not random,
but which have defined values at $t=0$. Due to these assumptions, this
environment is not at thermal equilibrium. The time-dependent frequency is written
in the form $ \omega (t) = \Omega + \delta \omega(t)$.

\begin{figure}[!ht]
\includegraphics[width=8.5cm]{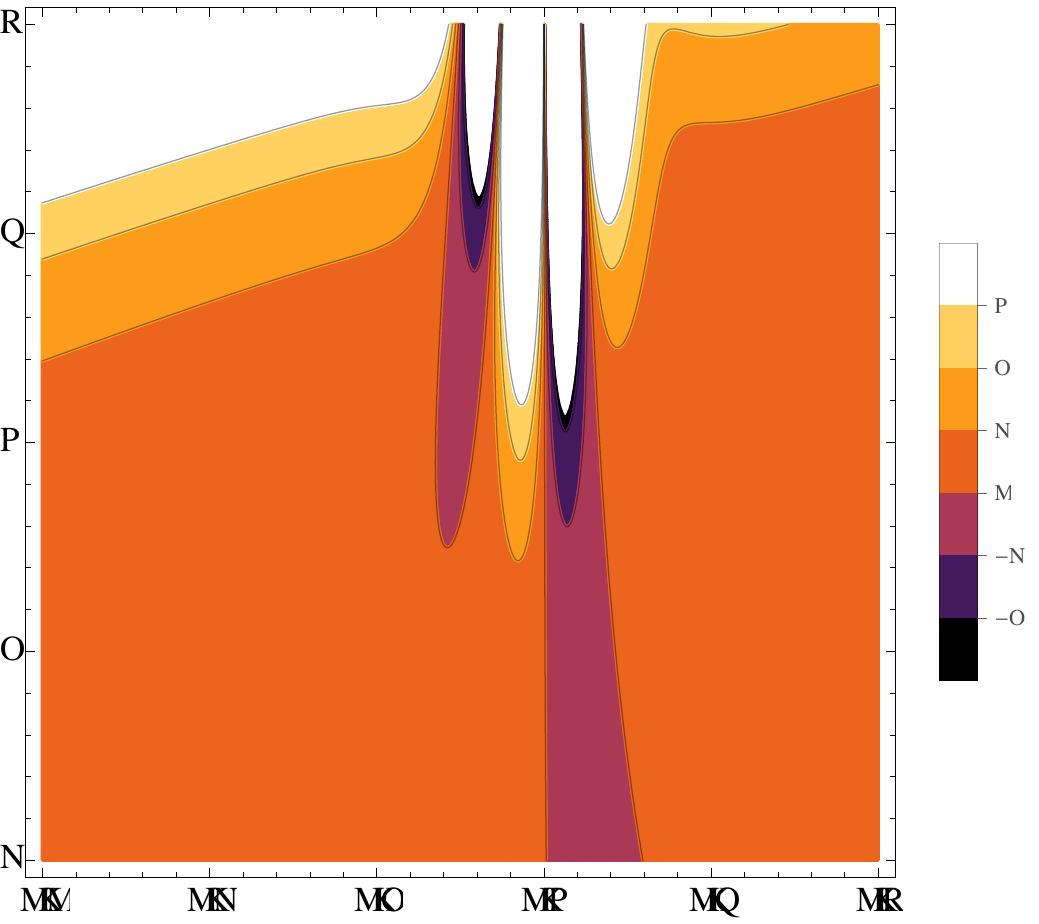}
\caption{(Color online) Density plot of the diffusion coefficient for the non-equilibrium environments. 
The horizontal axis represents the time evolution as $\Omega t$ and the vertical axis is the ohmicity $s$. The darkest areas are negative values
 of the diffusion coefficient. Parameters used:  
 $\Lambda=10\Omega$ , $\gamma_0=0.1$,
$\Omega\lambda=0.3$ and $d=2\Omega$.}
\label{difnq1}
\end{figure}

Following this approach,  it is easy to check that the off-diagonal terms of the reduced density matrix can be written as
\begin{eqnarray}
\rho_{\rm r_{01}}(t) &=& e^{-i \Omega t} < e^{- i \int_0^t \delta \omega(s)ds}> 
\rho_{\rm r_{01}}(0) \nonumber \\
&\equiv& e^{-i \Omega t} e^{-\tilde{{\cal F}}(t)} ~\rho_{\rm r_{01}}(0).
\end{eqnarray} 
 Here, we denote with $<...>$ the non-equilibrium average
over the non-stationary random bath and ${\tilde {\cal F}}(t)$ is defined as the decoherence
factor for the non-equilibrium environments, which reads,
\begin{eqnarray}
 {\tilde{\cal F}}(t) &=& \exp \left\{-\left(\gamma_0 \exp(-4 d t) (-1 + \exp(2 d t)\right. \right.
   \nonumber \\
&\times& \Gamma[1+s] ((1+4 (t-\lambda)^2 \Lambda^2)^{\frac{-(1+s)}{2}}  \nonumber \\
&\times& 
 \cos((1+s) \arctan(2 \Lambda (t-\lambda)))+ \cosh(2 d t)  \nonumber \\
 &+& \left.\left.\sinh(2 d t))
 \right)\right\} ,
 \label{factorneq}
\end{eqnarray}
where $d$ and $\lambda$ are parameters of the environment model.
By knowing the decoherence factor we can have a better insight into the decoherence process induced in the system by the presence
of a non-equilibrium environment.
\begin{figure}[!ht]
\includegraphics[width=8.5cm]{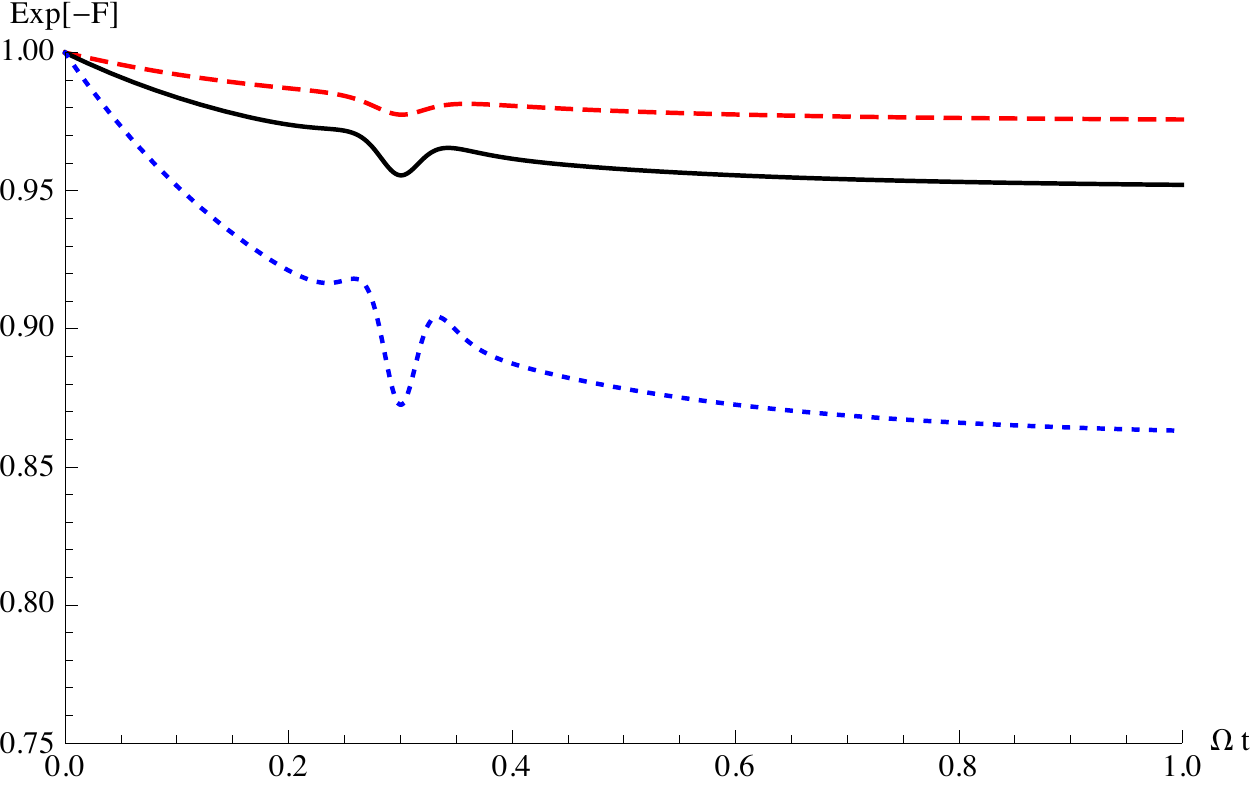}
\caption{(Color online) Decoherence factor for different values of the ohmicity parameter $s$ as function of time. We can see that the dip becomes more relevant and drastic for bigger values of $s$.
The red dashed line is for $s=1$, the black solid line is for $s=2$ and the blue dotted line for $s=3$. Parameters used:  
 $\Lambda=10\Omega$ , $\gamma_0=0.1$,
$\Omega\lambda=0.3$ and $d=2\Omega$.}
\label{fnq}
\end{figure}

In Fig.\ref{difnq1} we show a contour plot for the diffusion coefficients for
different forms of the ohmicity parameter $s$ in the vertical axis and different values of $\Omega t$ in the horizontal one. We can note that there is always a period
when the coefficient becomes negative, even for $s=1$. This means that memory effects are always present for this type of environments and the decoherence factors are always non-monotonic decreasing functions, allowing recoherence effects on the system of interest.
We can consider these environments to be non-Markovian for all values of $s$.
We can also note a clear hierarchy in the coefficients: the bigger the value of $s$, the 
more negative is the diffusion coefficient and the more decoherence (and recoherence) induced by the environment. 
This fact can be observed in Fig.\ref{fnq}, where we show the decoherence factor for different non-equilibrium environments. It can be seen that even though we have $s=1$, the behaviour is non-monotonic, quite different to the decoherence factor of a thermal equilibrium environment. In Fig.\ref{fnq} we present the decoherence factor for different values of the ohmicity parameter $s$ as function of time. The red dashed line corresponds to $s=1$, the black solid line is for $s=2$ and the blue dotted line shows the case $s=3$. In all cases, these non-equilibrium environments present a ``dip", which makes the behaviour of the decoherence factor non-monotonic. The strength and location of the ``dip" is determined by the other parameters of the model $\lambda$ and $d$. 

\section{Correction to the Geometric Phase}
In order to compute the geometric phase (GP) and note how it is corrected by the presence of the environment, 
we shall  briefly review  the way the GP
 can be computed for a system under
the influence of external conditions such as an external bath.
 In Ref. \cite{Tong}, a quantum kinematic
approach was proposed and the GP for a mixed state
under non-unitary evolution has been defined
as
\begin{eqnarray} 
\phi_G & = &
{\rm arg}\{\sum_k \sqrt{ \varepsilon_k (0) \varepsilon_k (\tau)}
\langle\Psi_k(0)|\Psi_k(\tau)\rangle \nonumber \\ &&\times e^{-\int_0^{\tau} dt \langle\Psi_k|
\frac{\partial}{\partial t}| {\Psi_k}\rangle}\}, 
\label{fasegeo}
\end{eqnarray}
where $\varepsilon_k(t)$ are the eigenvalues and
 $|\Psi_k\rangle$ the eigenstates of the reduced density matrix
$\rho_{\rm r}$ (obtained after tracing over the reservoir degrees
 of freedom). In the last definition, $\tau$ denotes a time
after the total system completes a cyclic evolution when it is
isolated from the environment. Taking into account the effect of the
environment, the system no longer undergoes a cyclic evolution.
However, we shall consider a quasi cyclic path ${\cal P}:~ t~
\epsilon~ [0,\tau]$, with $\tau= 2 \pi/ \Omega$ ($\Omega$ is the
system's characteristic frequency). When the system is open, the
original GP, i.e. the one that would have been obtained if the
system had been closed $\phi_u$, is modified. This means, in a
general case, that the phase can be interpreted as $\phi_G = \phi_u + \delta \phi$,
where $\delta \phi$ depends on the kind of environment coupled to
the main system \cite{pra, nos, pau, pra2, prl, pra3}.

We want to compute the GP acquired by the system for 
different forms of the reservoir spectrum, as function of the ohmicity parameter, 
which allows the description
of sub-Ohmic, Ohmic and super-Ohmic spectra. Particularly, 
we want to see how the unitary geometric phase for a two-level system $\phi_u= \pi(1-\cos\theta )$
is corrected  as a function
of the ohmicity, i.e. the parameter $s$ of the spectral density $I(\omega)$.
Assuming an initial quantum state of the system as 
\begin{equation}
 |\psi (0) > = \cos(\theta/2) |0> + ~\sin(\theta/2) |1>,
\end{equation}
its evolution at a later time $t$, is
\begin{equation}
 |\psi (t) > = e^{-i \Omega t} \cos(\theta_+(t)) |0> + ~\sin(\theta_+(t)) |1>,
\end{equation}
where $ \cos(\theta_+(t))$ (and $\sin(\theta_+(t))$) encodes diffusion induced on the subsystem due to the presence of the environment.  As explained in Ref.\cite{Tong},
the GP is obtained by computing eigenvectors and eigenvalues of the
reduced density matrix derived by using the state vector $| \psi(t)>$ and using Eq.(\ref{fasegeo}).


We shall consider the thermal equilibrium and non-equilibrium environments considered above. With the decoherence factors and the reduced density matrix computed analytically for each case, we can compute the GP and see how it is affected by the memory effects present 
in the different environments considered.

 \subsection{Thermal equilibrium environments}

 By considering the thermal equilibrium environments in the zero-T limit and using the corresponding decoherence factor (Eq.(\ref{factorexacto})), we can find an expression for the correction to the GP  
in terms of a series expression in the coupling constant 
\begin{eqnarray}
 \phi_G &=& \phi_u+  \gamma_0 \sin^2\theta \cos\theta 
\, \,  \Gamma(s-2)\Big\{ 2 \pi  (s-2) \nonumber \\
 &+&
  \Big(1+ \frac{4 \pi^2 \Lambda^2}{\Omega^2} \Big)^{-\frac{s}{2}}
 \Big[ 4 \pi \cos\Big(s \arctan\Big(\frac{2 \pi \Lambda}{\Omega}\Big)\Big) \nonumber \\
 &+& 
 (\frac{4 \pi^2 \Lambda}{\Omega}-\frac{\Omega}{\Lambda}) \sin \Big( s \arctan\Big(\frac{2 \pi \Lambda}{\Omega}\Big) \Big)\Big] \Big\}. \label{correq}
 \end{eqnarray}
 As it is expected, in the limit $s \rightarrow 1$ the correction to the phase approaches $\delta\phi \approx 4 \pi \gamma_0 \sin^2\theta \cos\theta \left( -1 + \log[2 \pi\Lambda/\Omega]\right)$.  
 In the limit $s \rightarrow 3$, the correction is given by $\delta\phi \approx 4 \pi \gamma_0 \sin^2\theta \cos\theta$. Both expressions agree with the corrections to the Ohmic 
($s = 1$) and supra-Ohmic ($s = 3$) geometric phases in the zero temperature limit found in Ref.\cite{pra}. As this result is derived for small values of the coupling constant $\gamma_0$, in the following we shall compute the exact GP numerically.\\

\begin{figure}[!ht]
\includegraphics[width=8.3cm]{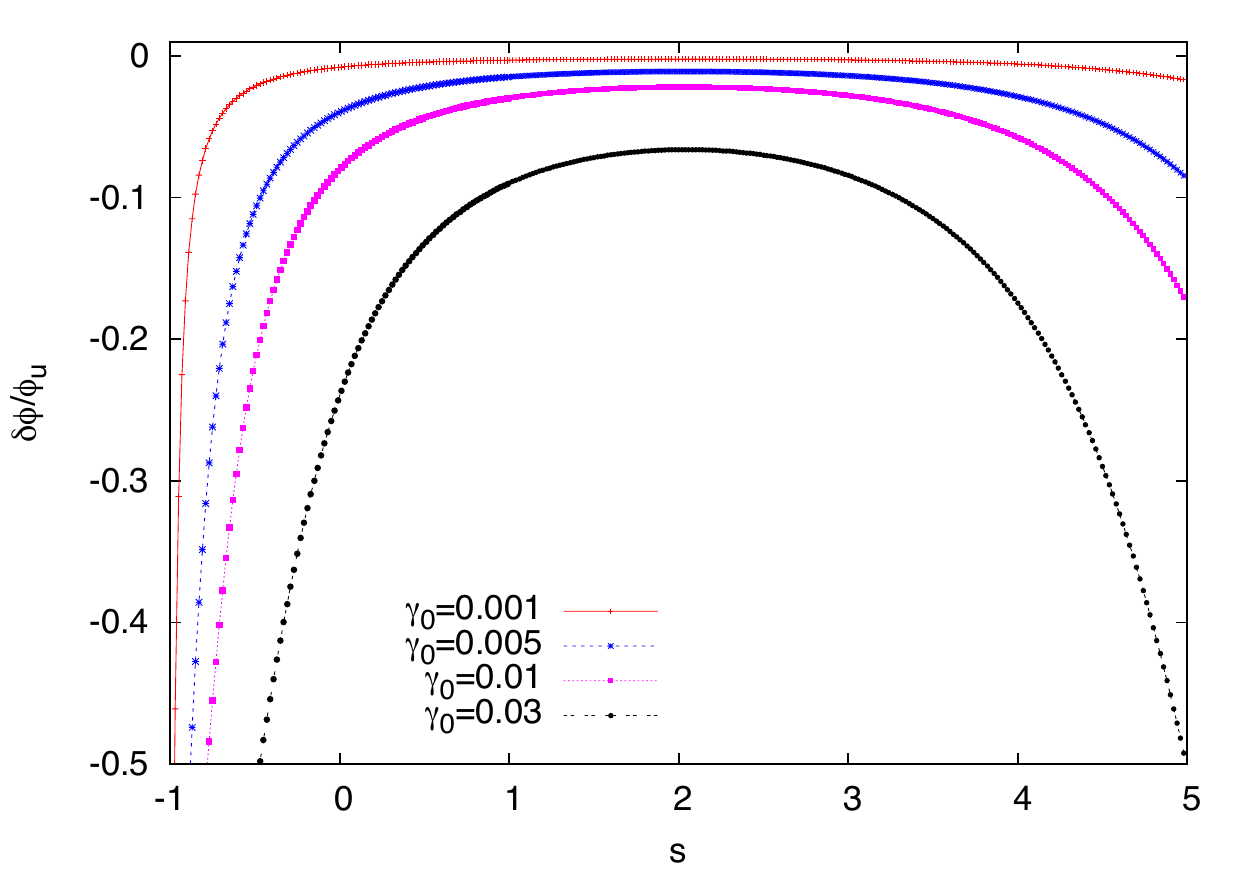}
\caption{(Color online) Correction to the geometric phase (in units of the unitary phase) as a function of the 
ohmicity parameter $s$. Red cross-solid line is for $\gamma_0=0.001$, blue asterisk-line is for $\gamma_0=0.005$,
magenta square-line is for $\gamma_0=0.01$ and black circle-line for $\gamma_0=0.03$ for zero T environments.
Parameters used: $\theta=\pi/3$, $\Lambda=10 \Omega$.}
\label{faseohm}
\end{figure}

In Fig.\ref{faseohm}, we show the correction induced by different environmental types on the  
geometric phase (normalised by the value of the unitary geometric phase $\phi_u$)
as a function of the ohmicity for different values of $\gamma_0$.
In this figure we can note that the geometric phase is very much destroyed when $s \rightarrow -1$. 
In that case, we are
considering the effect of a noise, similar to $1/f$, which is very harmful. This type of noise can be 
considered a sub-Ohmic
environment due to the fact that low frequencies are predominant. The correction to the phase grows 
as the effect of low frequency modes of the environment becomes more relevant. On the contrary,  as $s$  
starts increasing, we obtain the correction to the 
phase similar to the one of an Ohmic environment ($s=1$) at zero-T \cite{pra}. As can be expected, 
decoherence induced by this type of environment
is  low for a very weak coupling (less than $10\%$ for smaller values of $\gamma_0$). However, it becomes significant for bigger values of $\gamma_0$, for example for $\gamma_0=0.03$ in Fig.\ref{faseohm}, the correction is bigger than $20\%$. This agrees with the results in Ref.\cite{pra}:  more 
decoherence induced on the system, implies a bigger correction to the unitary geometric phase. 
 Near 
the {\it Ohmic region} of the parameters set, the correction depends mainly on $\gamma_0$, as expected, 
being negligible when the small value of $\gamma_0$ is such that is not able to 
destroy coherences in the system. 
For $1 < s < 2$ the correction to the phase is more insensitive to the ohmicity 
value, and it is evident that there is a change in behaviour for $s>2$ (it starts increasing), in agreement
with the onset of non-Markovianity. This shows that non-Markovian environments induce a bigger 
correction to the unitary geometric phase. As expected, 
this fact is enlarged for bigger couplings between system and environment, i.e. values of $\gamma_0$, 
since these imply a bigger decoherence effect. These corrections have not been studied previously, for example in Ref.\cite{pra}.
In conclusion, corrections to the phase are double. 
Firstly, we have the common hierarchy in the induced correction ruled by the coupling to
the environment: the bigger correction occurs with the bigger value of $\gamma_0$. 
Secondly, it is possible to see that the correction to the phase becomes more relevant 
when the ohmicity parameter surpasses its critical value. 
It becomes evident that the more non-Markovian the environment, the biggest correction to the 
geometric phase for the same coupling to the environment. This explains the behaviour observed  in Fig.\ref{faseohm}. 

\begin{figure}[!ht]
\includegraphics[width=8.3cm]{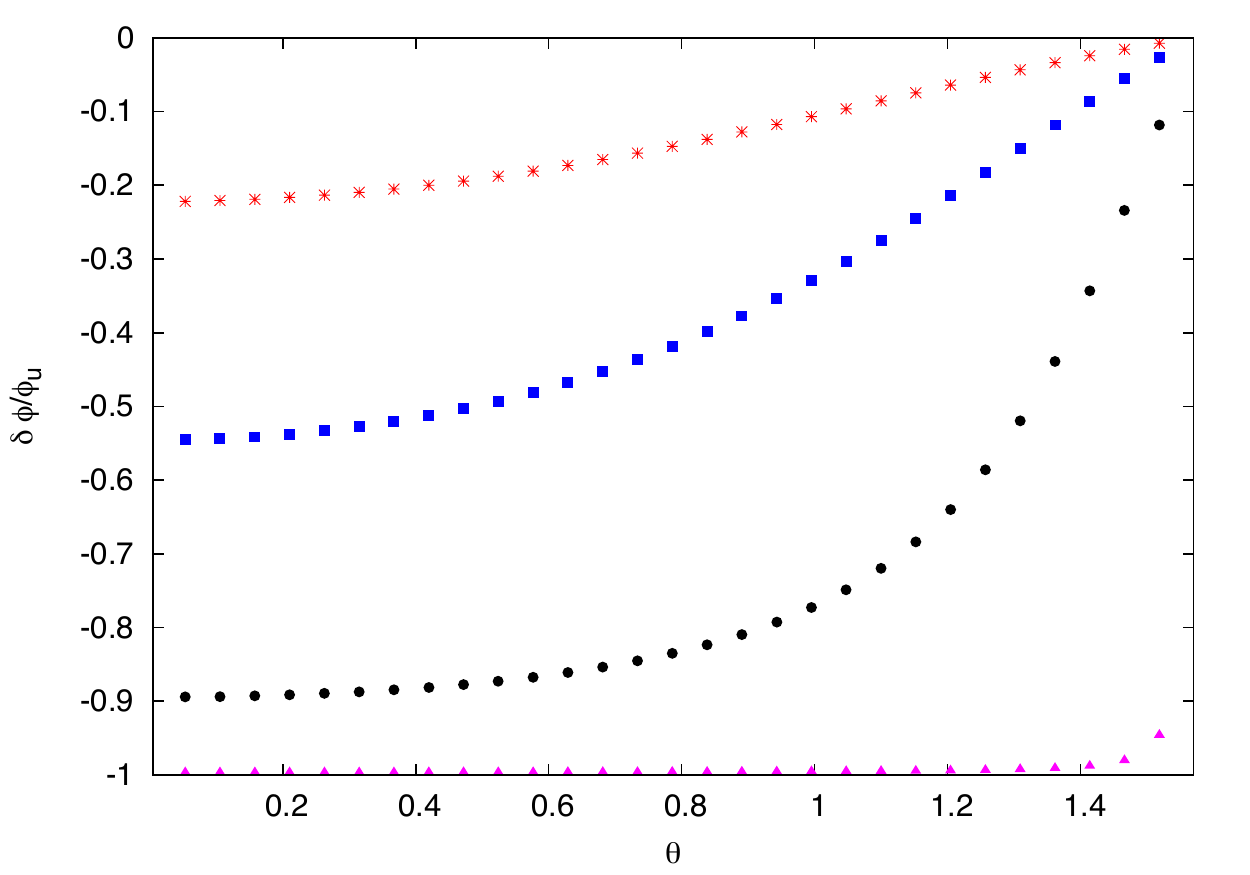}
\caption{(Color online) The correction to the GP as function of the angle 
that defines the state vector in the Bloch sphere ($\theta$) for Ohmic and supra-Ohmic environments in the zero
T limit. Red asterisks represent the Ohmic environment ($s=1$), while blue squares are for a supra-Ohmic environment of $s=2$ 
and black circles one of $s=2.5$. The triangles represent $s=3$. Parameters used: $\gamma_0=0.01$ and
$\Lambda=10 \Omega$.}
\label{comparison}
\end{figure}

In Fig.\ref{comparison} the normalised correction is plotted  for different environments at
 zero temperature. Red asterisks represent an Ohmic environment ($s=1$), while blue squares are for a 
 supra-Ohmic environment of $s=2$ and black circles one of $s=2.5$. 
 The triangles represent $s=3$. We can see that for $s>2$, the correction to the geometric phase is bigger and its behaviour is more drastic, in agreement with the onset of non-Markovianity expected to be for environments at zero temperature.

\subsection{Non-equilibrium environments}

Finally, we compute the correction to the GP for the non-equilibrium environments which have memory effects for all values of $s$. Therefore, we shall consider them as non-Markovian environments.

\begin{figure}[!ht]
\includegraphics[width=8.7cm]{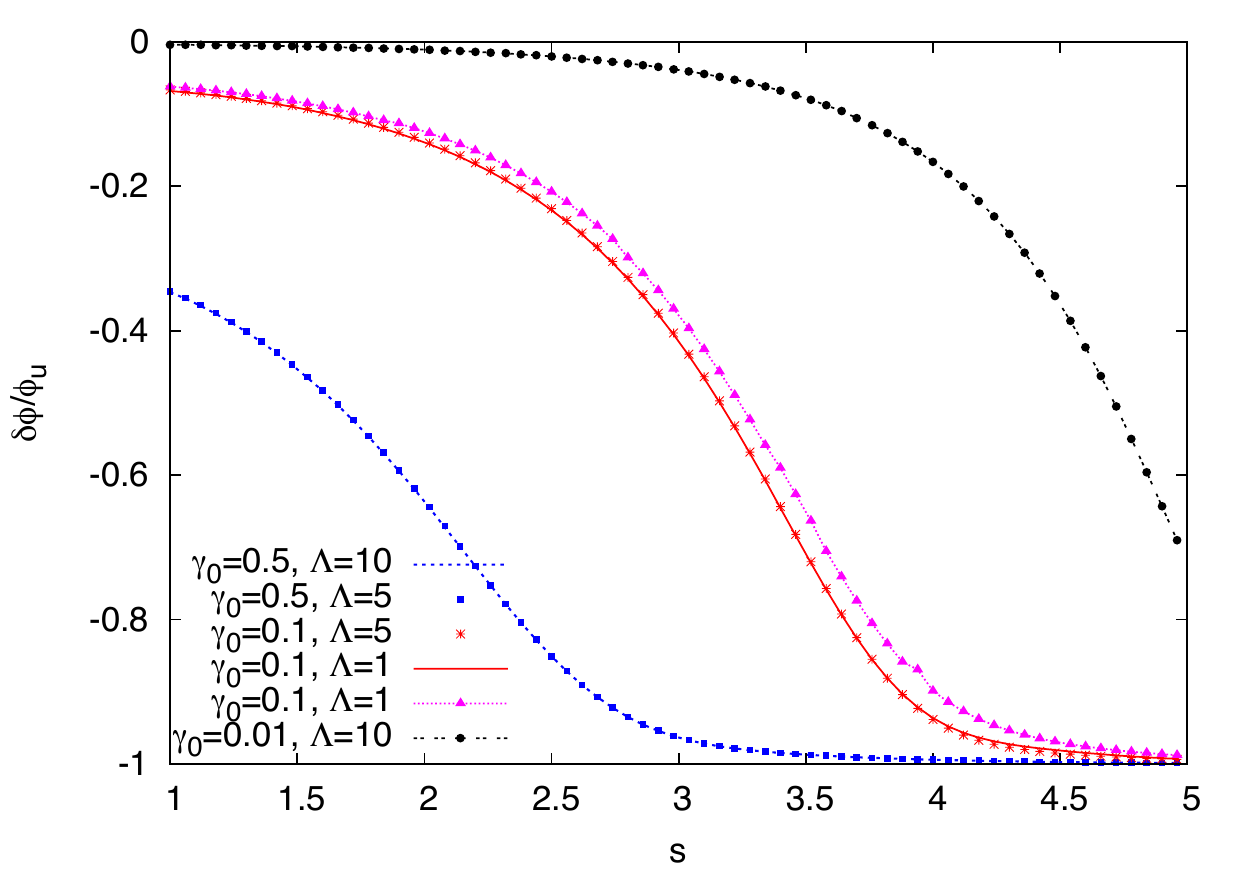}
\caption{(Color online) Correction to the unitary geometric phase for different values of the model's parameters.The blue dotted line is for $\gamma_0=0.5$ and $\Lambda=10\Omega$ while
the blue square dots is $\gamma_0=0.5$ and $\Lambda=5\Omega$. The red asterisks are for $\gamma_0=0.1$ and $\Lambda=5\Omega$,
while the solid red line is $\gamma_0=0.1$ and $\Lambda=1\Omega$ and the triangle dotted line 
$\gamma_0=0.1$ and $\Lambda=1\Omega$ but changing the parameters of $\lambda$ and $d$. Finally, the dotted black line is for
$\gamma_0=0.01$ and $\Lambda=10\Omega$. Parameters used:  $\Omega\lambda=0.5$, $d=1\Omega$. }
\label{GP}
\end{figure}

In Fig.\ref{GP}, we show the behaviour of the normalised correction to the unitary geometric phase  as a function of the 
ohmicity $s$. In this figure we can see that for smaller values of $s$ and particular values of $\lambda$ and
$d$ parameters (which determine a small dip in the decoherence factor coefficient as shown in Fig.\ref{fnq}), 
the non-equilibrium environment does not have a
big influence on the geometric phase, and the open geometric phase coincides with the unitary one. 
In Fig. \ref{GP}, we present different sets of values for the parameters of the environment's model. It is easy to note that the correction is mainly ruled by the value of the coupling constant: see, for example, the blue dotted line and the blue dots. Both lines correspond to the same value of $\gamma_0=0.5$ but different frequency cutoff $\Lambda$. The same occurs with the red line and the red asterisks with $\gamma_0=0.1$. The magenta triangles share the same value of $\gamma_0$ but the values of $\lambda$
and $d$ are interchanged with respect to the red line and asterisks. Finally, the black lines indicates smaller values of $\gamma_0$. When $s$ increases (even for the set of parameters  $\gamma_0$, $\lambda$ and $d$, for for which decoherence is negligible) the effect of the memory effects on the geometric phase 
of the non-equilibrium environment are stronger and 
the correction to the geometric phase results bigger. After this, the non-Markovian correction to the phase presents an abrupt slope, leading to bigger corrections compared to the equilibrium baths considered above.  By an analytical comparison of the perturbative 
Eqs. (\ref{correq}) and (\ref{corrnq}), it can be seen that  the correction induced by non-equilibrium environments is similar to that of the thermal ones for ohmicity parameters $-1<s<1$ (for equal coupling constant $\gamma_0$ and environment cutoff $\Lambda$). However, as $s$ increases, the correction induced by non-equilibrium environments gets bigger in comparison, becoming particularly important for $s>3$. This confirms the importance of the memory effects of the environment in the correction of the geometric phase, as already stated in 
 Refs. \cite{haikka} and \cite{Yi}, where authors studied the memory effects of the environment for  
different two-level systems. 

As in previous Section, we can estimate the correction to the unitary geometric phase, in an expansion 
in powers of the coupling between system and environment $\gamma_0$. In this case, the geometric phase can  be approximated, analytically as 
\begin{equation}
\phi_G  \approx \phi_u + \gamma_0  \, \Gamma[s + 1] \sin^2\theta \cos\theta \, , \label{corrnq}
\end{equation} where we have neglected ${\cal O}(\gamma_0 \Omega/\Lambda)$ terms. This approximate 
expression matches very well the low-$\gamma_0$ corrections given in Fig. \ref{GP}. This expression also 
gives the leading correction to the phase found in Ref. \cite{pra87} for $s = 1$ and $s= 3$. 
On general grounds, we can see that  
the non-equilibrium bath is more harmful
and induces a bigger correction on the geometric phase for all values of $\theta$ than the 
equilibrium environment. \\

\section{Conclusions}

The geometric phase of quantum states could have a potential application in holonomic quantum computation 
since the study of spin systems effectively allows us to contemplate
the design of a solid-state quantum computer. However, decoherence is the main
obstacle to overcome.  Furthermore, in most
cases of practical interest, quantum systems are subjected to many 
noise sources with different amplitudes and correlation times, corresponding
${\it de~facto}$ to a non-equilibrium environment.

We have computed the correction of the geometric phase under the presence of structured
reservoirs. We have defined the spectral density as function of the ohmicity parameter $s$. This is advantageous because it allows to study sub-Ohmic, Ohmic and supra-Ohmic environments in the same approach.
One could wonder if the correction to the geometric phase is due to the non-Markovianity of the environment or simply to a stronger effective interaction among the system and the environment.

Firstly, we have considered the  structured reservoirs to be composed of a set of harmonic oscillators at zero temperature. In the case of  thermal equilibrium environments, 
we have shown that for small values of the ohmicity $s \leq 2$, the hierarchy on
the correction of the unitary geometric phase is mainly due to the value of $\gamma_0$. This means: the stronger the coupling to the environment, the stronger
the correction induced on the geometric phase as expected. However, for $s>2$, even though the hierarchy is repeated, we
can see that the non-Markovianity has also an important role in the correction. We have checked this argument by considering
the decoherence factor for different values of $s$ and observing that the correction to the geometric phase is strongly influenced by this coefficient.

Secondly, we have  considered the structured environments to be modelled by a type of non-equilibrium environments.  The non-equilibrium feature is represented
by a non-stationary random function corresponding to the fluctuating transition
frequency between two quantum states coupled to the surroundings. 
We have shown that the diffusion coefficients have always negative values for same period of time for $s \geq 1$ which means a non-monotonic behaviour of the decoherence factors. This allows to consider these non-equilibrium environments as non-Markovian ones for all values of $s$.
In this framework, we have computed the non-unitary geometric phase
for the qubit in a quasi-cyclic evolution under the presence of these
particular non-equilibrium environments, both numerically and analytically.
When compared the correction of the geometric phase induced by these environments and the thermal ones, 
memory effects of the environments with random noise are more harmful and the correction 
induced is bigger particularly for values of $s \ge 3$.

Finally, these kind of environments could become
a proper experimental setup for the observation of the geometric phase. 
Our work on the assessment of the environmental effect on the geometric phase, especially in the non-Markovian regime (either for equilibrium or non-equilibrium environments) could be of great importance in using the geometric phases in two-level systems to implement the quantum gates.
The results presented in
this paper can also provide a clue to observe  the GP in a two-level system.

\end{document}